\documentclass[a4paper,11pt]{article}
\pdfoutput=1
\usepackage{jheppub} 
\usepackage{hyperref}
\usepackage{mathtools}
\usepackage{amsmath}
\usepackage{amsthm}
\usepackage{amssymb}
\usepackage{tikz}
\usepackage{graphicx}
\usepackage{slashed}
\usepackage{caption}
\usepackage{xcolor}
\usepackage{dsfont}
\usepackage{verbatim}
\usepackage{subfig}
\usepackage{enumitem}
\usepackage[T1]{fontenc} 
\newcommand\beq{\begin{equation}}
\newcommand\eeq{\end{equation}}
\newcommand\bal{\begin{aligned}}
\newcommand\eal{\end{aligned}}
\def\eqa{\begin{eqnarray}}

\title{Holographic boundary conformal field theory with $T\bar T$ deformation}

\author[a]{Zhi Wang,}
\author[b,1]{Feiyu Deng\note{Corresponding author.}}

\affiliation[a]{College of Physics, Nanjing University of Aeronautics and Astronautics, Nanjing 211106, China}
\affiliation[b]{Department of Physics and Center for Field Theory and Particle Physics, Fudan University, Shanghai 200433, China}

\emailAdd{zhiwang@nuaa.edu.cn}
\emailAdd{fydeng20@fudan.edu.cn}

\abstract{We propose a holographic dual of boundary conformal field theory (BCFT) with $T\bar T$ deformation, i.e. of $T\bar T$ BCFT. Our holographic proposal distinguishes two types of $T\bar T$ BCFTs, depending on whether the $T\bar T$ deformation deforms the boundary. For the boundary-deformed case, we find that boundary entropy serves as an effective measure to quantify the impact of boundary deformation. In this scenario, we calculate the energy spectrum for the $T\bar T$ BCFT within a finite interval to support the proposed dual. For the boundary-undeformed case, we calculate the entanglement entropy and Rényi entropy from both the field theory side and the gravity side, and find that they match.
}
\begin{document}

\maketitle

\flushbottom

\section{Introduction}
The AdS/CFT correspondence \cite{Maldacena:1997re, Gubser:1998bc, Witten:1998qj} is a powerful tool for studying strongly coupled conformal field theory (CFT) from a gravitational perspective. This correspondence has been extended to boundary conformal field theory (BCFT) by considering AdS gravity with an end-of-the-world (EOW) brane, known as AdS/BCFT duality~\cite{Takayanagi:2011zk, Fujita:2011fp}~\footnote{For recent progress on AdS/BCFT, please refer to \cite{Jensen:2013lxa, Miao:2017gyt, Almheiri:2018ijj, Cooper:2018cmb, Almheiri:2019hni, VanRaamsdonk:2020ydg, Rozali:2019day, Sully:2020pza, Geng:2020qvw, Bachas:2020yxv, Chen:2020tes, Raamsdonk:2020tin, Takayanagi:2020njm, Geng:2020fxl, Miyaji:2021ktr, Geng:2021iyq, Miyaji:2021lcq, Collier:2021ngi, Wang:2021xih, Geng:2021mic, Coccia:2021lpp, Grimaldi:2022suv, Suzuki:2022xwv, Hu:2022ymx, Kawamoto:2022etl, Izumi:2022opi, Miyaji:2022dna, Biswas:2022xfw, Kusuki:2022ozk, Kanda:2023zse, Kawamoto:2023wzj, Kawamoto:2023nki}.}.
 On the other hand, the correspondence has also been extended to field theories that flow away from the conformal fixed point, such as $T\bar T$-deformed CFTs~\cite{Zamolodchikov:2004ce, Smirnov:2016lqw, Cavaglia:2016oda}. Holographic duals of $T\bar T$-deformed CFTs is proposed to be AdS gravity within a finite cutoff region, at least in the pure gravity sector~\cite{McGough:2016lol, Guica:2019nzm}~\footnote{Recent studies on holographic $T\bar T$ CFT can be found in \cite{Donnelly:2018bef,Datta:2018thy,Chen:2018eqk,Hartman:2018tkw,Gorbenko:2018oov,Wang:2018jva,Gross:2019ach,Grieninger:2019zts,Lewkowycz:2019xse,Mazenc:2019cfg,Li:2020pwa,Belin:2020oib,Li:2020zjb,Caputa:2020lpa,Coleman:2021nor,Torroba:2022jrk,He:2023xnb,Apolo:2023vnm,Tian:2023fgf,Chang:2024voo,Babaei-Aghbolagh:2024hti}. }.
Then a natural question arises: can the AdS/CFT correspondence be generalized to handle CFTs with both a boundary and $T\bar T$ deformation? If the answer is yes, what is the bulk dual? The aim of this paper is to address this question, i.e., to study the gravity dual of holographic BCFTs with $T\bar T$ deformation~\footnote{For field theoretical studies on $T\bar T$ deformation with boundary,  see~\cite{Cavaglia:2016oda,Cardy:2018sdv,Babaro:2018cmq,Jiang:2021jbg,Brizio:2024doe}.}. Based on the bottom-up AdS/BCFT duality and the cutoff description of holographic $T\bar T$ CFT, we propose that the bulk dual of $T\bar T$ BCFT is AdS gravity enclosed by an EOW brane with Neumann boundary condition and a finite cutoff boundary with Dirichlet boundary condition. The EOW brane and the finite cutoff boundary intersect at the boundary of the $T\bar T$ BCFT~\footnote{In our previous work \cite{Deng:2023pjs}, a special case of holographic $T\bar T$ CFT with a boundary was obtained by taking the $Z_2$ quotient of the holographic $T\bar T$ CFT, corresponding to a $T\bar T$ BCFT with zero boundary entropy. In this work, we explore the bulk dual of a general holographic $T\bar T$ BCFT.}.

The combination of $T\bar T$ deformation and boundary introduces new possibilities in the correspondence. Depending on the background of the BCFT, the boundary of the BCFT can either undergo deformation or remain undeformed under the influence of the $T\bar T$ deformation. By distinguishing these two possibilities, we have discovered that their holographic duals exhibit differences. The key distinction lies in the behavior of the BCFT boundary within the AdS bulk during $T\bar T$ flows. In the case with a deformed boundary, the boundary undergoes movement in the AdS bulk as the $T\bar T$ operator evolves. This indicates that the boundary is deformed by the $T\bar T$ deformation and  we find the boundary entropy is a good quantity to quantify the deformation. We also check the agreement of energy spectrum of $T\bar T$ BCFT in a finite interval.
Conversely, in the case with an undeformed boundary, the BCFT boundary remains stationary in one position within the bulk, regardless of any variations in the $T\bar T$ operator. In this case, due to the property of the undeformed boundary, we calculate the vacuum entanglement entropy and the (refined) Rényi entropy from both the field theory side and the gravity side, and find that they match.

This paper is organized as follows. In Section~\ref{adsbcftttbar}, we review the main idea of AdS/BCFT duality and the cutoff description of holographic $T\bar T$ CFT. In Section~\ref{ttbcft}, we propose the bulk dual of holographic $T\bar T$ BCFT and classify it into the boundary-deformed and boundary-undeformed cases. In Section~\ref{Attbcft}, we calculate the boundary entropy and the energy spectrum for a finite interval in the boundary-deformed $T\bar T$ BCFT.
 In Section~\ref{Bttbcft}, we calculate the entanglement and Rényi entropy for the boundary-undeformed $T\bar T$ BCFT. In Section~\ref{cons}, we summarize the conclusions and discuss future directions.

\section{Review of holographic BCFT and holographic $T \bar T$ CFT}\label{adsbcftttbar}

In this section, we review the main idea of bottom-up  AdS/BCFT duality~\cite{Takayanagi:2011zk,Fujita:2011fp} and the cutoff description of holographic $T\bar T$ CFT~\cite{McGough:2016lol}. The bulk dual of holographic BCFT in $d$-dimensions is proposed to be AdS gravity in $d+1$ dimensions with an EOW brane $Q$ which intersects with the BCFT boundary. The embedding of the EOW brane $Q$ is determined by the Neumann boundary condition
\begin{equation}
      K_{a b}-\gamma_{a b} K=8\pi G_{ N}T_{ab}^Q\ \label{NBC}\ ,
\end{equation}
where $\gamma_{a b}$ is the induced metric on the EOW brane and $T_{ab}^Q$ is the stress energy tensor of the brane localized matter field. For example, the vacuum state of BCFT$_2$ is dual to pure AdS$_3$ with a constant tension EOW brane. By setting the boundary of BCFT$_2$ at the spatial origin and take the bulk to be AdS$_3$ in Gaussian normal coordinates
\begin{equation}
ds^2=d\rho^2+l^2\cosh^2{\frac{\rho}{l}}\left(\frac{-dt^2+du^2}{u^2}\right)\ ,
\end{equation}
the solution of Eq.~(\ref{NBC}) with a constant tension $T_{ab}^{Q}=T\gamma_{ab}$ tells us that the EOW brane $Q$ locates at $\rho=\rho_0$ slice. Thus the dual geometry is 
\beq\label{BCFTbulk}
ds^2=d\rho^2+l^2\cosh^2{\frac{\rho}{l}}\left(\frac{-dt^2+du^2}{u^2}\right)\ ,\quad \rho\leq\rho_0\ .
\eeq
The relation between $\rho_0$ and tension $T$ is
\begin{equation}\label{T}
    T=\frac{\tanh\frac{\rho_0}{l}}{l}\ .
\end{equation}
The brane tension $T$ is related to the BCFT boundary entropy as~\cite{Takayanagi:2011zk} 
\begin{equation}
S_{\mathrm{bdy}}=\frac{l\ \mathrm{arctanh} (Tl)}{4G_N}=\frac{\rho_0}{4 G_N}\ .
\end{equation}

Now we turn to cutoff description of holographic $T\bar T$ CFT. For a CFT$_2$ in flat space without boundary, one can iteratively deform the CFT by $T\bar T$ operator 
\beq\label{TTbar}
\frac{dS(\lambda)}{d\lambda}=-2\pi\int d^2x\; T\bar{T}\ ,
\eeq
where $\lambda$ is the $T\bar T$ deformation parameter and we choose the initial condition of this differential equation as $S(0)=S_{\mathrm{CFT}}$. The $T\bar T$ operator is defined as
\begin{equation}\label{tto}
    T \bar{T}=\frac{1}{8}\left(T^{\mu\nu} T_{\mu\nu}-\left(T_\mu^\mu\right)^2\right)\ ,
\end{equation}
where $T^{\mu\nu}=\frac{2}{\sqrt{-g}} \frac{\delta S_{\lambda}}{\delta g_{\mu\nu}}$ is the stress energy tensor of the $T\bar T$ CFT. The cutoff description of holographic $T\bar{T}$ CFT is given by AdS gravity with a Dirichlet cutoff boundary, where the $T\bar T$ CFT resides~\cite{McGough:2016lol,Hartman:2018tkw}. For example, the vacuum state of $T\bar T$ CFT$_2$ in flat space $ds^2=-dt^2+dx^2$ is dual to pure AdS$_3$ in Poincare patch with a radial cut off at $z=z_c$
\begin{equation}\label{TTCFT}
ds^2=\frac{l^2}{z^2}(-dt^2+dz^2+dx^2)\ ,\quad z\geq z_c\ ,
\end{equation}
where the cut-off position $z_c$ and $T\bar T$ deformation coupling $\lambda$ is related by $\lambda=\frac{8 G_N}{l}z_c^2$.

\section{Holographic $T \bar T$ BCFT}\label{ttbcft}

\begin{figure}
	\centering
	\includegraphics[scale=0.4]{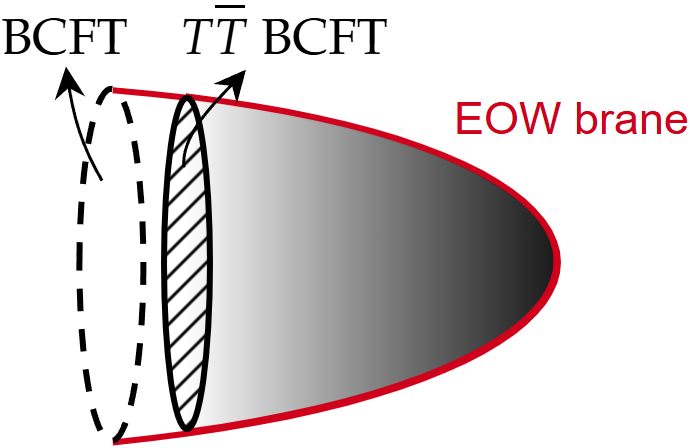}\\
	\caption{Bulk description of holographic $T\bar T$ BCFT. }\label{TTbarBCFT}
\end{figure}

Based on the spirit of bottom-up AdS/BCFT duality and cutoff description of holographic $T\bar T$ CFT, now we propose that the bulk dual of holographic $T\bar T$ BCFT is given by AdS gravity enclosed by an EOW brane with Neumann boundary condition and a finite cutoff boundary with Dirichlet boundary condition, they intersect with each other at the $T\bar T$ BCFT boundary. For a holographic BCFT deformed only by the $T\bar T$ operator, which is composed of BCFT bulk stress-energy tensor, the dual geometry is naturally constructed by introducing a finite cutoff boundary, similar to the case of holographic $T\bar T$ CFT. The $T\bar T$ BCFT then resides on this cutoff boundary.~\footnote{In this paper, we focus on considering BCFT deformed by BCFT bulk $T\bar T$ operator, which will not modify the location or geometry of the EOW brane. For $T\bar T$ BCFT with a boundary localized operator deformation, a natural way to obtain the dual geometry is to take the cutoff first and then resolve the Neumann boundary condition.} See Fig.~\ref{TTbarBCFT} for an illustration.

In general, a BCFT deformed by the $T\bar T$ operator in the BCFT bulk will induce a boundary flow. This induced flow is manifested in the dual geometry as the movement of the intersection point between the EOW brane and the cutoff boundary along the $T\bar T$ deformation. There is also the case where the bulk $T\bar T$ deformation does not deform the boundary. We categorize $T\bar T$ BCFTs based on whether the boundary is deformed or not. For simplicity, we refer to the boundary-deformed cases as Type A and the boundary-undeformed cases as Type B. Now, we introduce two typical examples of holographic Type A and Type B $T\bar T$ BCFTs, which are obtained by introducing the $T\bar T$ deformation into the AdS/BCFT setup. More evidences for the duality are provided in latter sections.

\subsection{Type A: boundary deformed }
\begin{figure}
	\centering
	\includegraphics[scale=0.4]{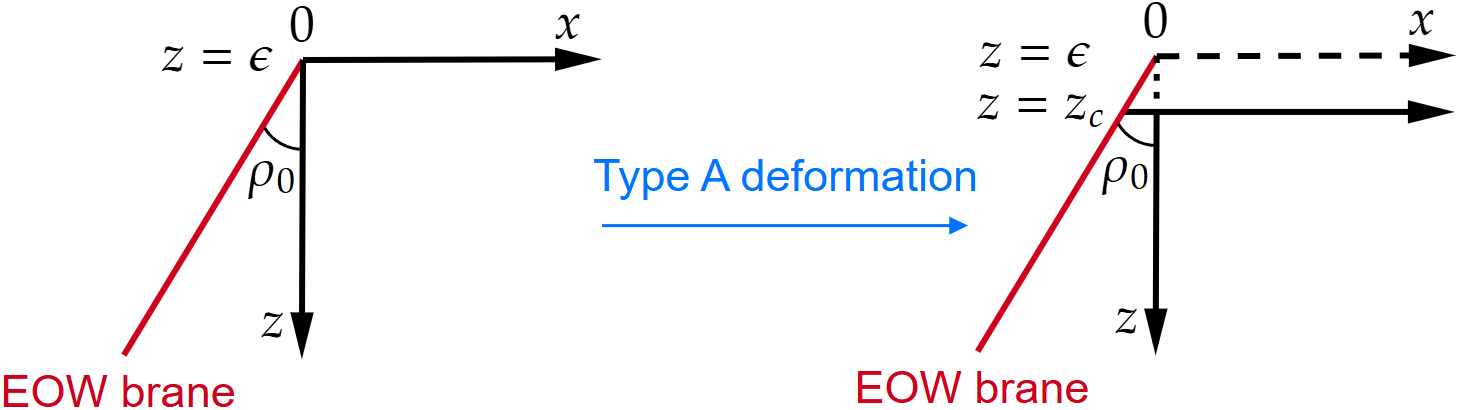}\\
	\caption{Holographic dual of Type A $T\bar{T}$ BCFT in half flat space. }\label{fig1} 
\end{figure}

A typical example for Type A $T\bar T$ BCFT is given by $T\bar T$ BCFT$_2$ in half flat space. According to our proposal, the bulk dual of the vacuum state for $T\bar T$ BCFT in half flat space is given by
\beq\label{TTBCFTbulk}
\bal
ds^2&=\frac{l^2}{z^2}(-dt^2+dz^2+dx^2) \\
&=d\rho^2+l^2\cosh^2{\frac{\rho}{l}}\left(\frac{-dt^2+du^2}{u^2}\right)\ ,\quad \rho\leq\rho_0\ ,\quad z\geq z_c\ .
\eal
\eeq
where $z=\frac{u}{\cosh\frac{\rho}{l}},x=-u\tanh\frac{\rho}{l}$. $z=z_c$ is the position of Dirichlet cutoff boundary and $\rho=\rho_0$ is the position of EOW brane. 
The bulk geometry Eq.~(\ref{TTBCFTbulk}) is the combination of Eq.~(\ref{BCFTbulk}) and Eq.~(\ref{TTCFT}), see Fig.~\ref{fig1} for an illustration. The holographic dictionary for deformation parameter $\lambda$ in $T\bar T$ BCFT is still given by $\lambda=\frac{8G_N}{l}z_c^2$. 
As shown in Fig.~\ref{fig1}, the $T\bar T$ deformation clearly shifts the boundary from $x=0$ to $x=-z_c\sinh\frac{\rho_0}{l}$, which indicates that the boundary is deformed.

\subsection{Type B: boundary undeformed}\label{Btt}

\begin{figure}
	\centering
	\includegraphics[scale=0.4]{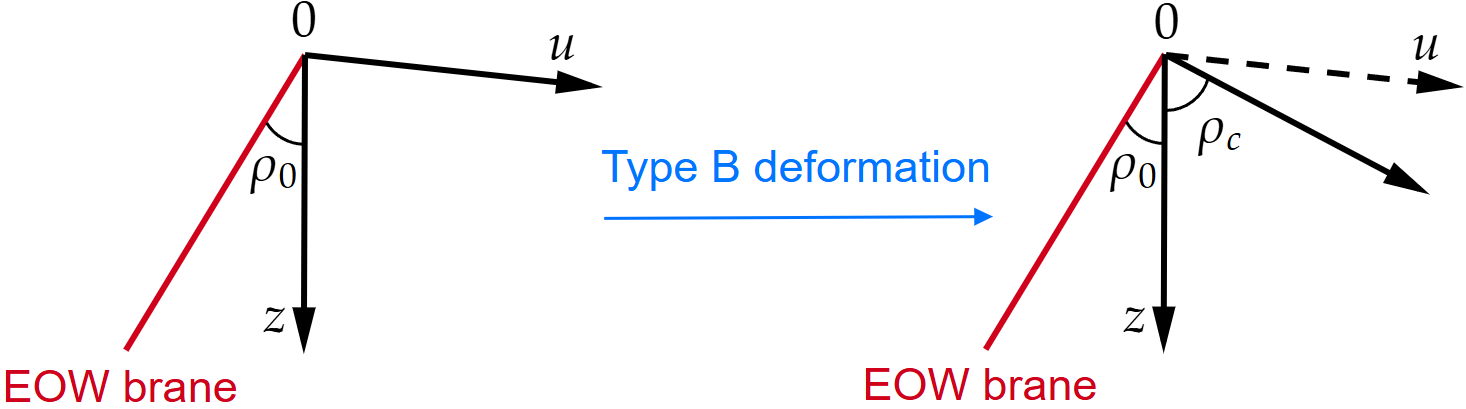}\\
	\caption{Holographic dual of Type B $T\bar{T}$ BCFT in AdS$_2$. }\label{fig2} 
\end{figure}

A typical example for Type B $T\bar T$ BCFT is given by $T\bar T$ BCFT$_2$ in AdS$_2$. In this case, the warped factor of background AdS$_2$ metric will fully suppress the bulk $T\bar T$ deformation, so the bulk $T\bar T$ deformation does not touch the boundary and will leave the BCFT boundary undeformed~\cite{Jiang:2019tcq,Brennan:2020dkw,Deng:2023pjs}. According to our proposal, the holographic dual of $T\bar T$ BCFT in AdS$_2$ is given by pure AdS$_3$ with a Dirichlet AdS$_2$ cutoff boundary and a Neumann EOW brane, they intersect with each other at asymptotic boundary. The bulk dual of the vacuum state for $T\bar T$ BCFT in AdS$_2$ is given by 
\begin{equation}
    d s^{2}
=d \rho^{2}+l^2\cosh ^{2} \frac{\rho}{l}\left(\frac{-d t^{2}+d u^{2}}{u^{2}}\right)\ ,\quad -\rho_c\leq\rho\leq\rho_0\ ,
\end{equation}
where $\rho=-\rho_c$ is Dirichlet cutoff boundary and $\rho=\rho_0$ is EOW brane, see Fig.~\ref{fig2} for an illustration.
The holographic dictionary is derived as follows. 

The extrinsic curvature on Dirichlet boundary is 
\begin{equation}
    K_{ab}=\frac{\tanh{\frac{\rho_c}{l}} }{l}h_{ab}=\frac{\tanh{\frac{\rho_c}{l}} }{l}\begin{bmatrix} 

    -\frac{l^2\cosh ^{2} \frac{\rho_c}{l}}{u^2} & 0 \\

      0 & \frac{l^2\cosh ^{2} \frac{\rho_c}{l}}{u^2}

      \end{bmatrix}\ , \qquad
\end{equation}
where $h_{ab}$ is the induced metric on $\rho=-\rho_c$ slice. Then the Brown-York tensor is computed as 
\begin{equation}
T_{ab}=\frac{1}{8 \pi G_N}\left(K_{ab}-K h_{ab}+\frac{1}{l} h_{ab}\right)
\end{equation}
and the result is 
\begin{equation}\label{BY}
    T_{ab}=-\frac{1}{8 \pi G_N}\frac{1-\tanh{\frac{\rho_c}{l}}}{l}\begin{bmatrix} 

    -\frac{l^2\cosh ^{2} \frac{\rho_c}{l}}{u^2} & 0 \\

      0 & \frac{l^2\cosh ^{2} \frac{\rho_c}{l}}{u^2}

      \end{bmatrix}\ .
\end{equation}
By using (\ref{BY}), one can rewrite $T_a^a$ in terms of $T \bar{T}$ operator and the result is 
\begin{equation}\label{TF}
    T_a^a=\frac{1}{8\pi G_N l}\frac{1}{\cosh^2{\frac{\rho_c}{l}}}-32\pi G_N l T\Bar{T}\ .
\end{equation}
 One can directly identify (\ref{TF}) with the $T\bar{T}$ deformed CFT trace flow equation, living in the curved spacetime
 \begin{equation}
     T_a^a=-\frac{c}{24 \pi} \mathcal{R}\left[h\right]-4 \pi \lambda T \bar{T}\ ,
 \end{equation}
 where the first term comes from trace anomaly and $\mathcal{R}\left[h\right]=-\frac{2}{l^2\cosh^2{\frac{\rho_c}{l}}}$ is the Ricci scalar computed by using $h_{ab}$. Thus one can determine the holographic dictionary to be
 \begin{equation}\label{dic21}
         c=\frac{3l}{2G_N}\ ,\quad\lambda=8G_Nl\ .
 \end{equation}
To see the cutoff dependence of deformation parameter, we can observe the trace flow equation under background metric $\gamma^{ab}=\cosh^2{\frac{\rho_c}{l}}h^{ab}$ and it is
\begin{equation}
     T_a^a=\frac{1}{8\pi G_N l}-\frac{32\pi G_N l}{\cosh^2{\frac{\rho_c}{l}}} T\bar{T}\ .
\end{equation}
Then we compare it with the trace flow equation of CFT living in curved spacetime with metric $\gamma^{ab}$ and we get 
\begin{equation}\label{dic22}
    \lambda=\frac{8G_Nl}{\cosh^2{\frac{\rho_c}{l}}}\ .
\end{equation}

\section{Boundary entropy and energy spectrum in Type A}\label{Attbcft}
In this section, we focus on Type A $T\bar T$ BCFT. First, we calculate the holographic boundary entropy of Type A $T\bar T$ BCFT using the disk partition function and find that it agrees with the boundary entropy extracted from holographic entanglement entropy. We also find that the boundary entropy can be used to quantify the boundary deformation for Type A $T\bar T$ BCFT. To provide further evidence for the proposed bulk dual of Type A $T\bar T$ BCFT, we calculate the energy spectrum for $T\bar T$ BCFT in a finite interval and find that the gravity theory result agrees with the field theory result.

\subsection{Boundary entropy}
\begin{figure}
	\centering
	\includegraphics[scale=0.4]{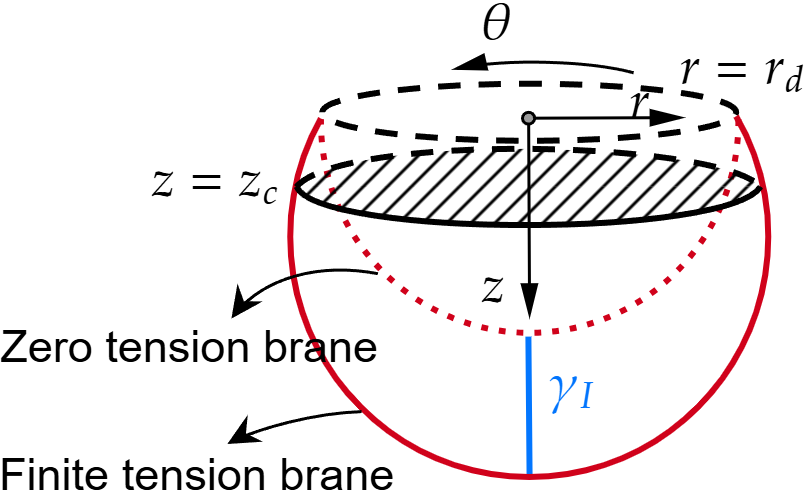}\\
	\caption{We choose BCFT zero tension brane as the reference brane in computing boundary entropy. The field theory degrees of freedom on the cutoff disk, located between the BCFT zero tension and finite tension brane, is treated as boundary degrees of freedom.}\label{TTbardisk2}
\end{figure}

\begin{figure}
	\centering
	\includegraphics[scale=0.4]{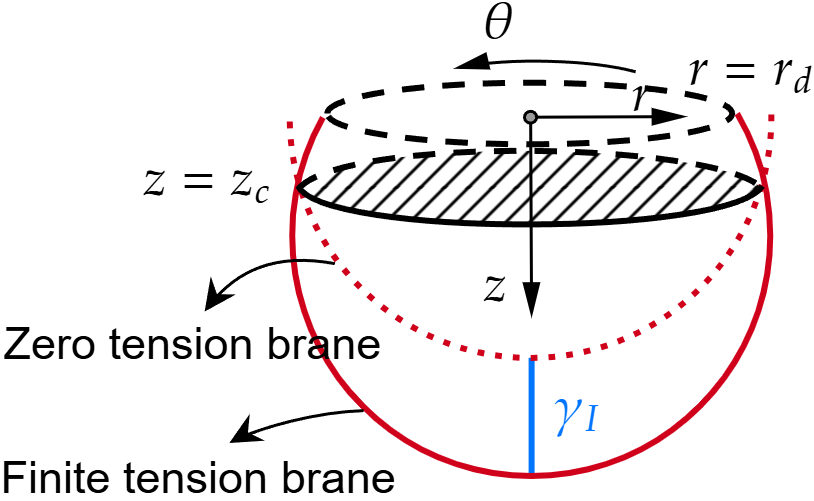}\\
	\caption{We choose a new zero tension brane, which intersects with the $T\bar T$ BCFT boundary, as a reference brane for computing the boundary entropy. The field theory degrees of freedom on the cutoff disk, located between the BCFT zero tension and finite tension brane, is treated as bulk degrees of freedom.
}\label{TTbardisk1}
\end{figure}

Now we compute the boundary entropy of a Type A holographic $T\bar T$ BCFT by considering $T\bar T$ BCFT in a disk. The boundary entropy is given by the amplitude~\cite{Affleck:1991tk}
\beq\label{bdp}
S_{\text{bdy}}^{\mathrm{disk}}=\log g=\log \langle 0|B\rangle\ ,
\eeq
where $|B\rangle$ is a general boundary state~\footnote{Since the $T\bar T$ deformation continuously deforms the spectrum and states of a given CFT, we consider this definition of boundary entropy to be equally well-defined for $T\bar T$ BCFT.
 }.
In holographic set up we can compute boundary entropy from bulk on-shell action~\cite{Takayanagi:2011zk}. According to our proposal, the dual geometry of the $T\bar T$ BCFT in a disk is given by
\beq
ds_{\mathrm{bulk}}^2=\frac{l^2}{z^2}\left(dz^2+dr^2+r^2d\theta^2\right)\ ,\quad z\geq z_c\ 
\eeq
with a spherical EOW brane
\beq
r^2+\left(z-r_d\sinh\frac{\rho_0}{l}\right)^2=r_d^2\cosh^2\frac{\rho_0}{l}\ ,
\eeq
where $r_d$ is the disk radius of the BCFT on the asymptotic boundary.
The brane induced metric is
\beq
ds_{\mathrm{brane}}^2=\frac{l^2}{z^2(r_c^2-z^2+2r_d z\sinh\frac{\rho_0}{l})}\left[r_d^2\cosh^2\frac{\rho_0}{l}dz^2+\left(r_d^2-z^2+2r_d z\sinh\frac{\rho_0}{l}\right)^2d\theta^2\right]\ ,
\eeq
and the extrinsic curvature of the brane is 
\beq
\bal
K&=-\frac{z(r)^2\left(\left(2r+z(r)z'(r)\right)\left(1+z'(r)^2\right)+rz(r)z''(r)\right)}{lrz(r)^2(1+z'(r)^2)^{\frac{3}{2}}}\\
&=\frac{2}{l}\tanh\frac{\rho_0}{l}\\
&=2T\ .
\eal
\eeq
The on-shell action of the bulk bounded by a finite tension brane is~\footnote{Note that when we calculate entanglement entropy by bulk on-shell action, we should not add a counter term to cancel the Weyl anomaly, which is different to the calculation of renormalized on-shell action~\cite{Caputa:2020lpa,Li:2020zjb,Apolo:2023vnm}. The reason is that entanglement entropy in quantum field theory is always divergent, and the Weyl anomaly will contribute to the entanglement entropy~\cite{Tian:2023fgf}.}
\beq
\bal
I_E=&I_{\mathrm{bulk}}+I_{\mathrm{brane}}+I_{\mathrm{ct}}\\
=&-\frac{1}{16\pi G}\int dx^3\sqrt{g}\left(R+\frac{2}{l^2}\right)-\frac{1}{8\pi G}\int d^2x\sqrt{h}(K-T)-\frac{1}{8\pi G}\int d^2x\sqrt{h}(K-\frac{1}{l})\\
=&-\frac{1}{16\pi G}\int_{z_c}^{r_d(\cosh\frac{\rho_0}{l}+\sinh\frac{\rho_0}{l})}dz\int_0^{\sqrt{r_d^2\cosh^2\frac{\rho_0}{l}-(z-r_d\sinh\frac{\rho_0}{l})^2}}dr \frac{-4}{l^2}\frac{l^3}{z^3}2\pi r\\
&-\frac{1}{8\pi G}\int_{z_c}^{r_d(\cosh\frac{\rho_0}{l}+\sinh\frac{\rho_0}{l})}dz\frac{1}{l}\tanh\frac{\rho_0}{l}2\pi \frac{l^2}{z^2}r_d\cosh\frac{\rho_0}{l}\\
&-\frac{1}{8\pi G}\int_{0}^{\sqrt{r_d^2\cosh^2\frac{\rho_0}{l}-(z_c-r_d\sinh\frac{\rho_0}{l})^2}}dr\frac{1}{l} 2\pi r \frac{l^2}{z_c^2}\\
=&-\frac{l}{4G}\left(\frac{\rho_0}{l}+\log \frac{r_d}{z_c}\right)
\ .
\eal
\eeq
As in Ref.~\cite{Takayanagi:2011zk}, we assume that the boundary entropy of the boundary state dual to a zero tension brane is zero, meaning that we need to subtract the on-shell action within a reference zero tension brane to calculate the holographic boundary entropy. For $T\bar T$ BCFT in a disk, the boundary of the disk will shift in the AdS bulk along the $T\bar T$ deformation. Then we have two choices for the reference zero tension brane. The first choice is the zero tension brane of the BCFT, as shown in Fig.~\ref{TTbardisk2}. This choice implies that we treat the field theory degrees of freedom on the cutoff disk, located between the BCFT zero tension and finite tension brane, as boundary degrees of freedom. Here, the boundary is defined as the intersection of the cutoff disk and the zero tension brane.
The second choice is a new zero tension brane which intersects with the $T\bar T$ BCFT boundary, as shown in Fig.~\ref{TTbardisk1}. This choice implies we regard the field theory degrees of freedom on the cutoff disk, located between the BCFT zero tension and finite tension brane, as bulk degrees of freedom. Now we calculate the boundary entropy separately for each of these two choices.

For the first choice, the on-shell action within the BCFT zero tension brane is
\beq
I_E|_{\rho_0\rightarrow{0}}=-\frac{l}{4G}\log \frac{r_d}{z_c}\ ,
\eeq
Then the boundary entropy is given by disk partition function as
\beq
S_{\mathrm{bdy}}^{\mathrm{disk}}=-(I_E-I_E|_{\rho_0\rightarrow{0}})=\frac{\rho_0}{4G}\ .
\eeq

For a BCFT, the boundary entropy can also be extracted from entanglement entropy as
\beq\label{eebdy}
S_{\mathrm{bdy}}=S^{\mathrm{BCFT}}([0,L])-\frac{1}{2}S^{\mathrm{CFT}}([-L,L])\ ,
\eeq
where $S^{\mathrm{BCFT}}([0,L])$ is the vacuum entanglement entropy of interval $[0,L]$ for a 
BCFT in half flat space and $S^{\mathrm{CFT}}([-L,L])$ is the vacuum entanglement entropy of the CFT in the full flat space. According to AdS/BCFT, we find that this implies the holographic boundary entropy is simply given by the minimal surface $\gamma_I$ between the zero-tension brane and the finite-tension brane, i.e.
\beq
S^{\mathrm{RT}}_{\mathrm{bdy}}=\frac{\mathrm{Area}(\gamma_I)}{4G}\ .
\eeq 
In our case of $T\bar T$ BCFT in a disk, the minimal surface $\gamma_I$ between the finite-tension brane and the new zero-tension brane is determined by rotational symmetry, as shown in Fig.~\ref{TTbardisk2}. The boundary entropy computed by $\gamma_I$ is
\beq
S_{\mathrm{bdy}}^{\mathrm{RT}}=\frac{1}{4G}\int_{r_d}^{r_d(\cosh\frac{\rho_0}{l}+\sinh\frac{\rho_0}{l})}dz \frac{l}{z}=\frac{\rho_0}{4G}\ .
\eeq
We find that this result agrees with the one obtained from the disk partition function. Notice that in this case, the boundary entropy is the same as BCFT boundary entropy without $T\bar T$ deformation. This indicates in Type A $T\bar T$ BCFT, if we regard the degree of freedom in the extended interval between BCFT zero tension brane and finite tension brane as boundary degrees of freedom, then the total boundary entropy will be conserved along $T\bar T$ deformation. 

To quantify the effect of $T\bar T$ deformation on the boundary, we take the second choice. In this case, we need to solve the new zero tension brane which intersects with the $T\bar T$ BCFT boundary, as shown in Fig.~\ref{TTbardisk1}. 
The on-shell action of the bulk bounded by the new zero tension brane is
\beq
I_E|_{\rho_0\rightarrow{0},r_d\rightarrow{\sqrt{r_d(r_d+2z_c\sinh\frac{\rho_0}{l})}}}=
-\frac{l}{8G}\log\frac{r_d(r_d+2z_c\sinh\frac{\rho_0}{l})}{z_c^2}\ .
\eeq
Then the boundary entropy is obtained as
\beq
\bal
S_{\mathrm{bdy}}^{\mathrm{disk}}&=-(I_E-I_E|_{\rho_0\rightarrow{0},r_d\rightarrow{\sqrt{r_d(r_d+2z_c\sinh\frac{\rho_0}{l})}}})\\
&=\frac{\rho_0}{4G}-\frac{l}{8G}\log(1+\frac{2z_c}{r_d}\sinh\frac{\rho_0}{l})\ .
\eal
\eeq
In this case, we observe that the boundary entropy is no longer constant; it depends on the $T\bar T$ deformation parameter, or equivalently, on $z_c$. This result clearly indicates that the boundary of the BCFT is deformed, and we can use boundary entropy to quantify the amount of boundary deformation. The boundary entropy can also be computed by the minimal surface $\gamma_I$ between finite tension brane and the new zero tension brane, as shown in Fig.~\ref{TTbardisk1}. The result is
\beq
S_{\mathrm{bdy}}^{\mathrm{RT}}=\frac{1}{4G}\int_{\sqrt{r_d(r_d+2z_c\sinh\frac{\rho_0}{l})}}^{r_d(\cosh\frac{\rho_0}{l}+\sinh\frac{\rho_0}{l})}dz \frac{l}{z}=\frac{\rho_0}{4G}-\frac{l}{8G}\log(1+\frac{2z_c}{r_d}\sinh\frac{\rho_0}{l})\ .
\eeq
We can see it agrees with the result obtained by disk partition function.

\subsection{Energy spectrum}
Now we compute the energy spectrum for a $T\bar T$ BCFT in a finite interval. Let us start from the bulk dual of a high temperature CFT state, which is the BTZ black hole 
\begin{equation}\label{btz}
    d s^2=l^2\left(\frac{f(z)}{z^2} d \tau^2+\frac{d z^2}{f(z) z^2}+\frac{d x^2}{z^2}\right)\ ,
\end{equation}
where $f(z)=1-\left(\frac{z}{z_H}\right)^2$, $\tau \sim \tau+2 \pi z_H$, the temperature of the black hole is $T_{\mathrm{CFT}}=\frac{1}{2 \pi z_H}$.
We begin by first considering the addition of a conformal boundary to the CFT at $(x,z)=(0,0)$ and solving for the EOW brane. By assuming constant tension and solving the Neumann boundary condition, the trajectory of the EOW brane is determined to be~\cite{Takayanagi:2011zk} 
\begin{equation}
    x=-z_H \cdot \operatorname{arcsinh}\left(\frac{l T z}{z_H \sqrt{1-l^2 T^2}}\right)\ .
\end{equation}

\begin{figure}
	\centering
	\includegraphics[scale=0.4]{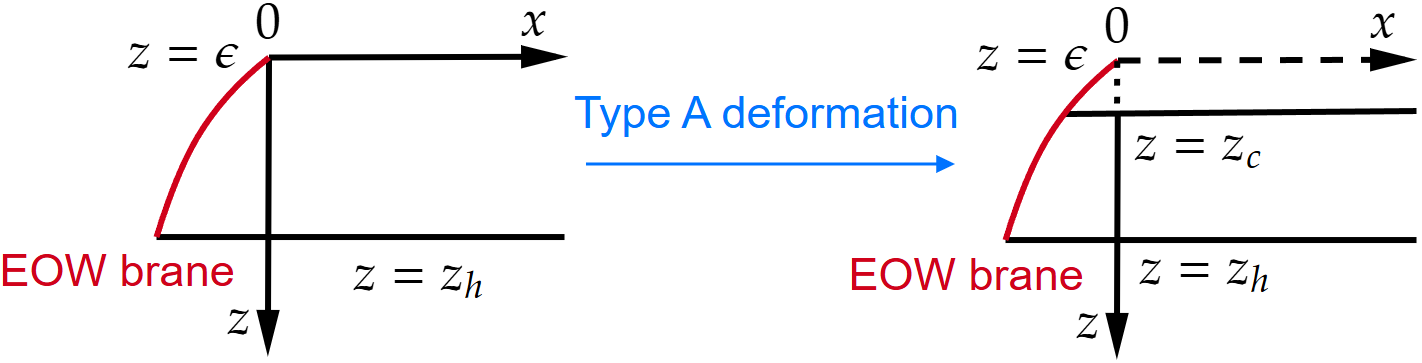}\\
	\caption{Holographic dual of thermal state for a Type A $T\bar T$ BCFT in half flat space. }\label{AB}
\end{figure}

Consequently, after the Type A deformation, the field theory exists on the cutoff boundary $z=z_c$, and the boundary of the field theory shifts from $(x,z)=(0,0)$ to $(x,z)=(-z_H \cdot \operatorname{arcsinh}\left(\frac{l T z}{z_H \sqrt{1-l^2 T^2}}\right),z_c)$. The corresponding bulk picture is illustrated in F.G.\ref{AB}.

\begin{figure}
	\centering
	\includegraphics[scale=0.35]{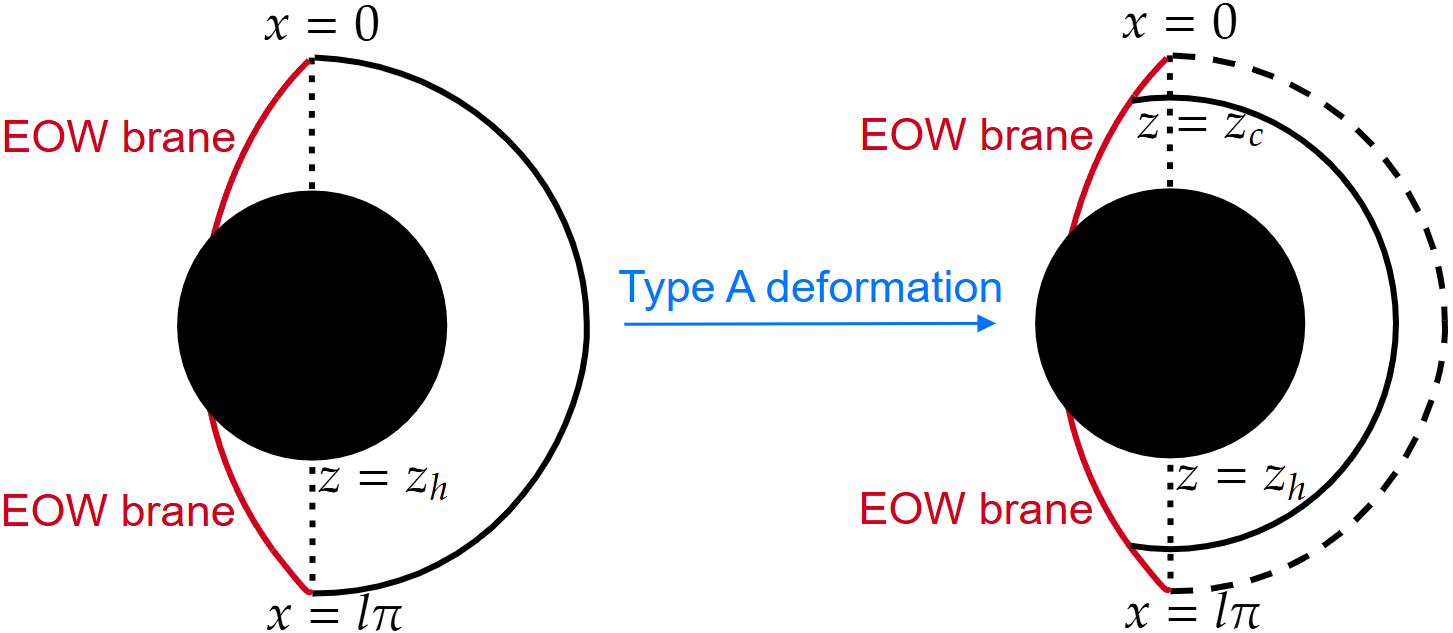}\\
	\caption{Holographic dual of thermal state for a $T\bar T$ BCFT in a finite interval.}\label{ES}
\end{figure}

To compute the energy spectrum of Type A $T\bar T$ BCFT in finite interval, we consider the spatial coordinate $x$ to be periodic $x\sim x+2\pi l$ and set $x=l \theta$ where $-\pi \leqslant \theta \leqslant \pi$.
We add two boundaries to the original CFT. One sits at $x=0$, the other locates at $x=l\pi$, as shown in Fig.~\ref{ES}. We aim to match the energy spectrum for $T\bar T$ BCFT bulk.
We first focus on deriving the energy spectrum from the gravity side. To start, we recall the gravity calculation of energy spectrum without the boundary. The energy can be straightforwardly obtained by first calculating the energy density from the Brown-York tensor of the cutoff boundary
\begin{equation}
    e=u^i u^j T_{i j}=\frac{u^i u^j }{8\pi G}\left(K_{i j}-K h_{i j}+h_{i j}\right)\ ,
\end{equation}
where $u^i$ is the unit vector that is normal to the constant $\tau$ slice of the cutoff boundary
\begin{equation} u^{\tau}=\frac{z_c}{l\sqrt{f(z_c)}}\ ,\quad
u^{\theta}=0\ .
\end{equation}
Then we get 
\begin{equation}\begin{split}
    e=\frac{1}{8\pi G l}\left(1-\sqrt{f(z_c)}\right)=\frac{1}{8\pi G l}\left(1-\sqrt{1-\frac{z_c^2}{z_H^2}}\right)=\frac{1}{8\pi G l}\left(1-\sqrt{1-\frac{8 G_N M z_c^2}{l^2}}\right)\ ,
\end{split}
\end{equation}
where we have used $z_H^2=\frac{l^2}{8G_N M}$.
The energy can be obtained by subsequently integrating over the spatial region. The result is
\begin{equation}
    E=\int_0^{2\pi} d \theta \sqrt{g_{\theta \theta}} e=\frac{l}{4G_Nz_c}\left(1-\sqrt{1-\frac{8 G_N M z_c^2}{l^2}}\right)\ .
\end{equation}
By multiplying the energy with the proper length $L=\int_0^{2\pi} d \theta \sqrt{g_{\theta \theta}}=\frac{2\pi l^2}{z_c}$, one can further get the dimensionless quantity
\begin{equation}
    \mathcal{E}=E L=\frac{\pi l^3}{2G_Nz_c^2}\left(1-\sqrt{1-\frac{8 G_N M z_c^2}{l^2}}\right)\ .
\end{equation}

Next we consider the energy spectrum after the EOW brane is introduced. We note that the energy density remains unchanged, as the boundary does not affect the Brown-York tensor; only the spatial region changes due to the introduction of boundaries.
Thus we can get the energy with boundaries as 
\begin{equation}\begin{split}
    E&=\int^{\pi+z_H/l \cdot \operatorname{arcsinh}\left(\frac{l T z}{z_H \sqrt{1-l^2 T^2}}\right)}_{-z_H \cdot \operatorname{arcsinh}\left(\frac{l T z}{z_H \sqrt{1-l^2 T^2}}\right)} d \theta \frac{l^2}{z_c}e 
\\&=\int^{\pi+z_H/l \cdot \operatorname{arcsinh}\left(\frac{l T z}{z_H \sqrt{1-l^2 T^2}}\right)}_{-z_H \cdot \operatorname{arcsinh}\left(\frac{l T z}{z_H \sqrt{1-l^2 T^2}}\right)}  \frac{l}{8 \pi G_N z_c}\left[1-\sqrt{1-\frac{8 G_N M z_c^2}{l^2}}\right]d \theta\\
    &=\frac{l}{8 \pi G_N z_c}\left[\pi+2z_H/l \cdot \operatorname{arcsinh}\left(\frac{l T z}{z_H \sqrt{1-l^2 T^2}}\right)\right]\cdot\left[1-\sqrt{1-\frac{8 G_N M z_c^2}{l^2}}\right]\ .
\end{split}\end{equation}
The bulk picture tells us that the boundaries only truncate the spatial circle while preserving the energy density. This suggests we can quantify their impact by recognizing that the effect is simply to rescale the spatial length. The rate of rescaling is
\begin{equation}\begin{split}
    f(\lambda)\equiv\frac{1}{2\pi}\left[\pi+\frac{2z_H}{ l} \operatorname{arcsinh}\left(\frac{l T z_c}{z_H \sqrt{1-l^2 T^2}}\right)\right]=\frac{1}{2}+\frac{z_H}{\pi l} \operatorname{arcsinh}\left(\frac{l T \sqrt{\frac{\lambda l}{8 G_N}}}{z_H \sqrt{1-l^2 T^2}}\right)\ .
\end{split}\end{equation}
Finally we can multiply the energy with the spatial length to get the dimensionless quantity
\begin{equation}\begin{split}
    \mathcal{E}&=E\cdot
    \int^{\pi+z_H/l \cdot \operatorname{arcsinh}\left(\frac{l T z}{z_H \sqrt{1-l^2 T^2}}\right)}_{-z_H \cdot \operatorname{arcsinh}\left(\frac{l T z}{z_H \sqrt{1-l^2 T^2}}\right)} d \theta \frac{l^2}{z_c}\\
    &=\frac{l^3}{8 \pi G_N z_c^2}\left[\pi+2z_H/l \cdot \operatorname{arcsinh}\left(\frac{l T z}{z_H \sqrt{1-l^2 T^2}}\right)\right]^2\cdot\left[1-\sqrt{1-\frac{8 G_N M z_c^2}{l^2}}\right]\ .
\end{split}
\end{equation}

Now let us compute the energy spectrum from the field theory side. Without boundaries, the spatial length of field theory is $L=2\pi l$. After adding boundaries, the spatial length of deformed theory is $L_b=2\pi l f(\lambda)=Lf(\lambda)$, where $f(\lambda)$ is the rescaling rate. The stress energy tensor of $T\bar T$ BCFT in finite interval  have zero momentum, i.e. $T_{\tau x}=T_{x \tau}=0$~\cite{Cardy:2018sdv,Brizio:2024doe}. Then we can separately compute the expectation value of $T\bar T$ operator as
\begin{equation}\begin{split}
    \langle n|T \bar{T}| n\rangle&=\frac{1}{8}\langle n|T| n\rangle\langle n|\bar T| n\rangle-\frac{1}{8}\langle n|\Theta| n\rangle\langle n|\Theta| n\rangle\\
    &=-\frac{1}{4}\left(\left\langle n\left|T_{\tau \tau}\right| n\right\rangle\left\langle n\left|T_{x x}\right| n\right\rangle\right)\ ,
\end{split}\end{equation}
and expectation value of components of stress tensor is
\begin{equation}
    \begin{split}
\left\langle n\left|T_{\tau \tau}\right| n\right\rangle &=\frac{E_n}{L_b}=\frac{E_n}{L f(\lambda)}\ , \\
\left\langle n\left|T_{x x}\right| n\right\rangle &=\frac{\partial E_n}{\partial L_b}=\frac{\partial E_n}{\partial L} \cdot \frac{1}{f(\lambda)}\ .
\end{split}
\end{equation}
From the definition of $T\bar T$ deformation
\begin{equation}
    S_{T\bar T}=2 \pi \lambda \int d^2 x T \bar{T}\ ,
\end{equation}
we have 
\begin{equation}\label{H}
    H_{T\bar T}=2 \pi \lambda\int^{Lf(\lambda)}_{0} d \theta \ T \bar{T}\ ,
\end{equation}
thus
\begin{equation}\label{pH}
\begin{split}
\frac{\partial}{\partial \lambda}\langle H_{T\bar T}\rangle  &=2 \pi Lf(\lambda)\langle T \bar{T}\rangle+2 \pi L\lambda f^{'}(\lambda)\langle T \bar{T}\rangle|_{Lf(\lambda)}
 \\
 &=2 \pi L f(\lambda)\langle T \bar{T}\rangle+\frac{f^{'}(\lambda)}{f(\lambda)}\langle H_{T\bar T}\rangle\ ,
\end{split}
\end{equation}
where we have used $\langle H_{T\bar T}\rangle=2 \pi \lambda Lf(\lambda)\langle T \bar{T}\rangle|_{Lf(\lambda)}=2 \pi \lambda Lf(\lambda)\langle T \bar{T}\rangle$ from (\ref{H}) in the second equality\footnote{The $T\bar T$ operator in differential equation (\ref{pH}) is the operator before deformation, it does not depend on the deformation parameter $\lambda$ and the position. Thus we have $\langle T \bar{T}\rangle=\langle T \bar{T}\rangle|_{Lf(\lambda)}$.}. Then we get
\begin{equation}
    \frac{\partial E_n}{\partial \lambda}=2 \pi L f(\lambda)\langle T \bar{T}\rangle+\frac{f^{'}(\lambda)}{f(\lambda)}E_n\ .
\end{equation}
The final equation becomes
\begin{equation}
    2 \frac{\partial E_n}{\partial \lambda}-\frac{2f^{'}(\lambda)}{f(\lambda)}E_n+\frac{ \pi E_n}{f(\lambda)} \frac{\partial E_n}{\partial L}=0\ .
\end{equation}
The solution is 
\begin{equation}
    E_n=\frac{L f(\lambda)}{\pi\lambda}\left[1-\sqrt{1-\frac{4\pi^2\lambda M_n}{L^2}}\right]\ ,
\end{equation}
where $M_n=\Delta_n+\bar{\Delta}_n-\frac{c}{12}$. 
By multiplying $E_n$ with $Lf(\lambda)$, one can therfore get the dimensionless quantity
\begin{equation}
    \mathcal{E}_N=\frac{L^2 f(\lambda)^2}{\pi\lambda}\left[1-\sqrt{1-\frac{4\pi^2\lambda M_n}{L^2}}\right]\ ,
\end{equation}
This result agrees with the gravity outcome under the identification $M_n=M\cdot l$ and the holographic dictionary $\lambda=\frac{8 G}{l} z_c^2$.

\section{Entanglement entropy and Rényi entropy in Type B}\label{Bttbcft}
In this section, we focus on Type B holographic $T\bar T$ BCFT and provide evidence for its bulk dual by calculating entanglement quantities. We continue to consider $T\bar T$ BCFT in AdS$_2$ as introduced in Section~\ref{Btt}. Since the boundary is undeformed in Type B $T\bar T$ BCFT, the contribution of boundary to entanglement entropy is always the BCFT boundary entropy. Below we follow the method of Ref.~\cite{Donnelly:2018bef} to calculate the nonperturbative entanglement entropy for a special interval of $T\bar T$ BCFT in AdS$_2$. We also compute Rényi entropy by using replica trick and cosmic brane. We find that the field theory result matches with the gravity theory result.

\subsection{Entanglement entropy}
\begin{figure}
	\centering
\includegraphics[scale=0.4]{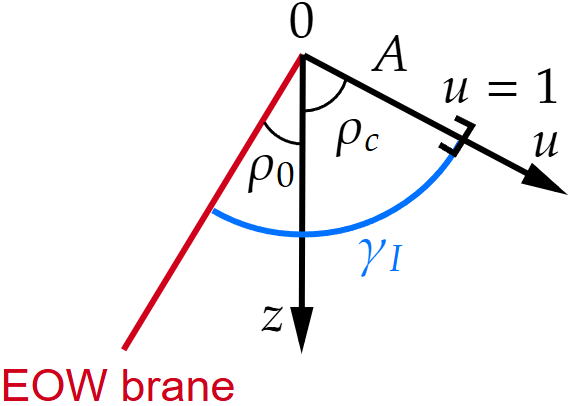}\\
	\caption{Ryu-Takayanagi (RT) surface for the interval  $u\in[0,1]$ in Type B $T\bar{T}$ BCFT. }\label{RT2}
\end{figure}

\begin{figure}
	\centering
	\includegraphics[scale=0.4]{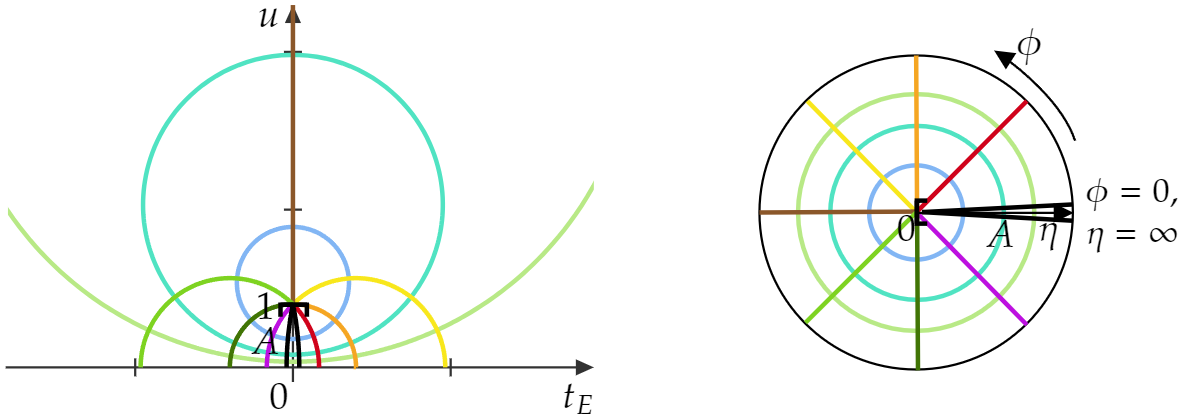}\\
	\caption{\label{PHH} Coordinate transformation between Euclidean Poincare AdS$_2$ coordinate $(t_E,u)$ in left figure and Hartle-Hawking coordinate $(\eta,\phi)$ in right figure. For interval $A$, the $n$-replica space is obtained by cyclic gluing $n$ copies of $\mathbb{H}^2$ along the cut $\phi=0$.}
\end{figure}

Now we compute the entanglement entropy (EE) for a $T\bar T$ BCFT in AdS$_2$. We consider the EE of an interval $A$ with $u \in [0,1]$ and $t = 0$ in the cutoff boundary, as shown in Fig.~\ref{RT2}.
This interval is special as it becomes a radius of the Euclidean AdS$_2$, making the replica geometry simple. To be specific, the Euclidean AdS$_2$ is represented as a hyperbolic disk
\begin{equation}
\begin{split}
ds_{\mathbb{H}^2}^2
        =r^2\frac{dt_E^2+du^2}{u^2}=r^2\left(d\eta^2+\sinh^2\eta d\phi^2\right)\ ,\\
\end{split}
\end{equation}
where 
\begin{equation}
    t_E=\frac{\sinh\eta\sin\phi}{\cosh\eta+\cos\phi\sinh\eta}\;\;\;,\;\;\;
    u=\frac{1}{\cosh\eta+\cos\phi\sinh\eta}\ ,
\end{equation}
and $r=l\cosh\frac{\rho_c}{l}$ is the AdS radius of the cutoff boundary. This transformation is shown in Fig.~\ref{PHH}. The metric of replica space along interval $A$ is 
\beq\label{adsin}
ds^2_{\mathbb{H}_n^2}=r^2\left(d\eta^2+n^2\sinh^2\eta d\phi^2\right) \ .
\eeq

Using replica trick, the EE of $A$ is obtained as
\begin{equation}\label{ee2}
\begin{split}
    S(A)=\left(1-n\frac{d}{dn}\right)\log Z_n\bigg|_{n=1}=\left(1-\frac{r}{2}\frac{d}{dr}\right)\log Z\ ,
\end{split}
\end{equation}
where in the second line we utilize two equations involving variation of the $T\bar T$ partition function with respect to $n$
\beq
\frac{d \log Z_n}{dn}\bigg|_{n=1}=-\frac{1}{2}\int d^2x\sqrt{h}\;\langle T^a_a\rangle\ ,
\eeq
and with respect to $r$
\beq
\frac{d}{d r}\log Z
        =-\frac{1}{r}\int d^2x \sqrt{h}\langle T^a_a\rangle\ .
\eeq
Then according to Eq.~(\ref{ee2}), in order to obtain $S(A)$, we need to evaluate the partition function $\log Z$ for $T\bar T$ deformed CFT in $\mathbb{H}^2$. This can be accomplished due to the fact that $\mathbb{H}^2$ is a maximally symmetric space.

Initially, in a maximally symmetric space, the stress tensor is proportional to the metric, i.e. $\langle T_{ab}\rangle=\alpha h_{ab}$. By substituting this into the trace flow equation, we can determine $\alpha$ and obtain
\begin{equation}\label{st}
    \langle T_{ab}\rangle=\frac{1}{\pi\lambda}\left(1-\sqrt{1-\frac{c\lambda}{12 r^2}}\right)h_{ab}\;\;\;,\;\;\;\langle T_{a}^{a}\rangle=\frac{2}{\pi\lambda}\left(1-\sqrt{1-\frac{c\lambda}{12 r^2}}\right)\ .
\end{equation}
Therefore, the variation of $\log Z$ with respect to $l$ for $\mathbb{H}^2$ is
\begin{equation}\label{dllogz}
    \frac{d}{d r}\log Z=\frac{4}{\lambda}\left(r-\sqrt{r^2-\frac{c \lambda}{12}}\right)\ .
\end{equation}
Note that we have another differential equation for $\log Z$ with respect to the deformation scale $\mu=\frac{1}{\sqrt{\lambda}}$
\begin{equation}\label{df}
\mu\partial \mu \log Z=-2\lambda\partial_{\lambda} \log Z=-\int d^2x \sqrt{h}\langle T^a_a\rangle=\frac{4r}{\lambda}\left(r-\sqrt{r^2-\frac{c \lambda}{12}}\right)\ .
\end{equation}
We can obtain $\log Z$ by integrating these two equations and imposing a proper boundary condition. We take the boundary condition to be
\begin{equation}
\begin{split}
\log Z|_{\rho_c=0}&=-I_E|_{\rho_c=0}=-\frac{1}{8\pi G_N}\int_{\rho=0}d^2x\sqrt{\gamma}\frac{1}{l}-I_{\mathrm{bdy}}=\frac{l}{4 G_N}-I_{\mathrm{bdy}}=\frac{c}{6}-I_{\mathrm{bdy}}\ ,
\end{split}
\end{equation}
where $I_{\mathrm{bdy}}$ is the on-shell action from $\rho=0$ to $\rho=\rho_0$, which is related to the boundary entropy as $I_{\mathrm{bdy}}=-S_{\mathrm{bdy}}=-\frac{\rho_0}{4G_N}$.
Using this boundary condition we obtain $\log Z$ 
\begin{equation}\label{pt2}
\log Z=\frac{c}{6} \log \left[\frac{r}{a}\left(1+\sqrt{1-\frac{c \lambda}{12 r^2}}\right)\right]-\frac{2 r^2}{\lambda} \sqrt{1-\frac{c \lambda}{12 r^2}}+\frac{2 r^2}{\lambda}+S_{\mathrm{bdy}}\ ,
\end{equation}
where $a=\sqrt{\frac{c\lambda}{12}}$ is a finite cutoff of the $T\bar T$ deformed theory. 
We can verify this boundary condition by evaluating the bulk Euclidean on-shell action and the result is
\begin{equation}\label{acb}
\begin{split}
        I_E&=-\frac{1}{16\pi G_N}(-2\pi) \int_{0}^{\rho_c}d\rho \;l^2\cosh^2\frac{\rho}{l}(-\frac{4}{l^2})\\
        &-\frac{1}{8\pi G_N}(-2\pi)l^2\cosh^2\frac{\rho_c}{l}(\frac{2}{l}\tanh\frac{\rho_c}{l}-\frac{1}{l})+I_{\mathrm{bdy}}\\
        &=-\frac{l}{8G_N}\left(1+e^{-\frac{2\rho_c}{l}}+\frac{2\rho_c}{l}\right)\ +I_{\mathrm{bdy}}\ .
\end{split}
\end{equation}
We find that it exactly equals to $-\log Z$ in Eq.~(\ref{pt2}) with $a=\sqrt{\frac{c\lambda}{12}}$ and Eq.~(\ref{dic21}).

Having obtained $\log Z$, the EE of $S(A)$ is given by Eq.~(\ref{ee2}) as
\begin{equation}
    S(A)
    =\frac{c}{6}\log\frac{r\left(1+\sqrt{1-\frac{c\lambda}{12 r^2}}\right)}{\sqrt{\frac{c\lambda}{12}}}+S_{\mathrm{bdy}}
    =\frac{c}{6}\frac{\rho_c}{l}+\frac{c}{6}\frac{\rho_0}{l}=\frac{\rho_c}{4G_N}+\frac{\rho_0}{4G_N} \ .
\end{equation}
The result matches with the holographic entanglement entropy given by $\gamma_I$ which we shown in Fig.~\ref{RT2}
\beq
S(A)=\frac{\mathrm{Area}(\gamma_I)}{4G_N}=\frac{1}{4G_N}\int_{-\rho_c}^{\rho_0}d\rho=\frac{\rho_c}{4G_N}+\frac{\rho_0}{4G_N}\ .
\eeq
It's worth noting that the result for zero boundary entropy case has been obtained in \cite{Deng:2023pjs}.
\subsection{Rényi entropy}
\begin{figure}
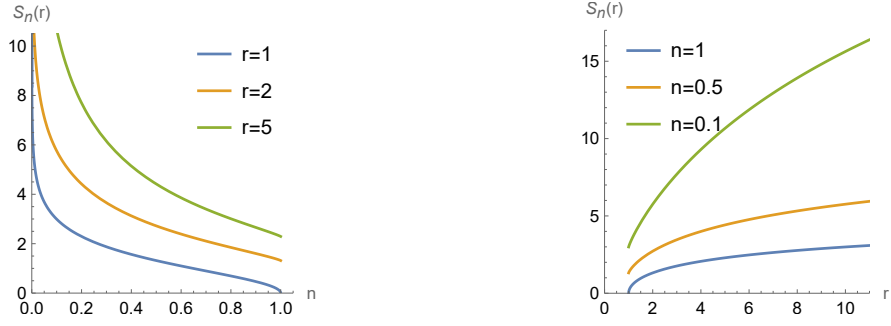

	\begin{minipage}[t]{0.49\linewidth}
		\centering
		\includegraphics[width=1.6in]{Sn.png}
	\end{minipage}
	\begin{minipage}[t]{0.49\linewidth}
		\centering
		\includegraphics[width=1.6in]{Sr.png}
	\end{minipage}
	\caption{Rényi entropy for different values of replica number $n\in (0,1]$ and different values of AdS radius $r$.}
	\label{Rényi}
\end{figure}

Now we compute (refined) Rényi entropy of the interval $A$ in the zero boundary entropy case. 
In the field theory side, the Rényi entropy can be calculated by using replica trick. 
The partition function in replica space (\ref{adsin}) satisfies 
\beq\label{lzn}
\frac{d\log Z_n}{dr}=-2\pi n r\int d\eta \sinh\eta\left(T^\eta_\eta+T^\phi_\phi\right)\ .
\eeq
The trace of stress energy tensor can be solved by combining the trace flow equation and conservation equation
\beq
UV=\frac{c\lambda}{24}\mathcal{R}[h]+1\ ,\quad
\partial_\eta U+\coth\eta (U-V)=0\ ,
\eeq
where $
U\equiv \pi \lambda T^\eta_\eta-1,  V\equiv \pi \lambda T^\phi_\phi-1
$. We set $T^\phi_\eta = 0$ and make the stress tensor independent of $\phi$ by utilizing the rotational symmetry along the $\phi$ direction. 
The result is given by
\beq
U(\eta)=-\sqrt{1-\frac{c \lambda}{12r^2}\left(1-\frac{1}{\sinh^2\eta}\frac{1-n^2}{n^2}\right)}\ ,
\eeq
The partition function can be obtained by integrating Eq. (\ref{lzn}) with a proper boundary condition. We imposed the boundary condition that when $n=1$, $\log Z_n$ going back to $\log Z$ that we have obtained in Eq.~(\ref{pt2}). Then $\log Z_n$ is obtained as
\beq
\log Z_n=
\frac{n}{6\lambda}\int d\eta \sinh\eta \left[-c\lambda \ \mathrm{arccoth} \frac{1}{-U(\eta)}-12r^2(1+U(\eta))\right]\ .
\eeq
The Rényi entropy is then obtained as
\beq
S_n=-n^2\frac{d}{dn}\left(\frac{\log Z_n}{n}\right)=-\frac{1}{4G}\frac{i r}{\sqrt{1-n^2}}\Pi \left[\frac{-n^2}{1-n^2};i\eta \Big|\frac{n^2(r^2-l^2)}{l^2(1-n^2)}\right]\ ,
\eeq
where the dictionary $c=\frac{3l}{2G},\lambda=8G l$ is used, and 
\beq
\Pi[p;\phi|q]\equiv \int_0^\phi d\theta\frac{1}{(1-p\sin^2\theta)\sqrt{1-q\sin^2\theta}}
\eeq
is the elliptic integral of the third kind. The plot of the Rényi entropy is shown in Fig.~\ref{Rényi}.

\begin{figure}
	\centering
	\includegraphics[scale=0.4]{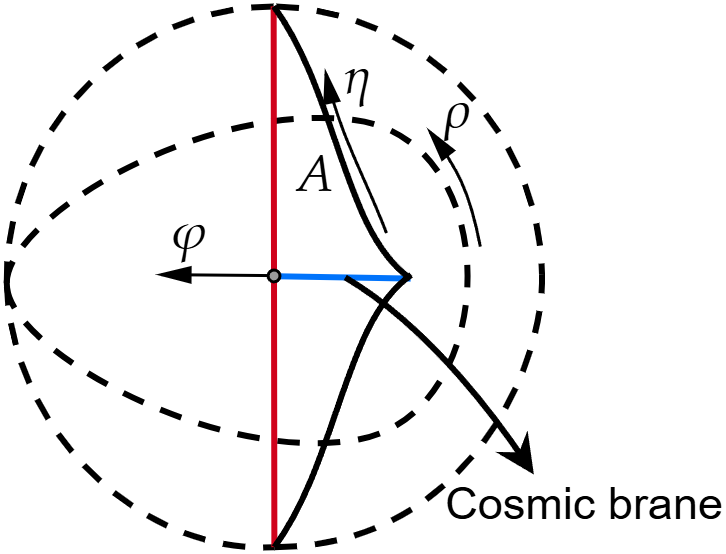}\\
	\caption{The cosmic brane is represented by the blue line, which lies along the axis of rotational symmetry.}\label{cosbra}
\end{figure}

In gravity side, the Rényi entropy is given by the area of cosmic brane. To find the configuration of the cosmic brane, we need to find an embedding of the replica metric into Euclidean AdS$_3$, as shown in Fig.~\ref{cosbra}. 
In Poincare coordinate of Euclidean AdS$_3$, the metric is given by 
\beq
ds^2=\frac{l^2}{z^2}(dz^2+d\rho^2+\rho^2d\phi^2)=l^2(d\varphi^2+e^{-2\varphi}(d\rho^2+\rho^2d\phi^2))\ ,
\eeq
where $z=e^\varphi$. We denote the embedding of $\mathbb{H}_n^2$ in Euclidean AdS$_3$ by $\varphi(\eta),\rho(\eta)$. By demanding that the induced metric is given by Eq.~(\ref{adsin}) , we obtain
\beq
\rho=e^\varphi\frac{r}{l}n\sinh\eta\ ,\quad \frac{r^2}{l^2}=(\frac{d\varphi}{d\eta})^2+e^{-\varphi}(\frac{d\rho}{d\eta})^2\ .
\eeq
From this we can get
\beq
\frac{d\varphi_{\pm}}{d\eta}=\pm \frac{r\sqrt{l^2(1-n^2)+n^2(r^2-l^2)\sinh^2\eta}}{l^2+n^2r^2\sinh^2\eta}\mp \frac{n^2r^2\cosh\eta\sinh\eta}{l^2+n^2r^2\sinh^2\eta}\ .
\eeq
The Rényi entropy is given by the area of the cosmic brane as
\beq
\bal
S_n&=\frac{1}{4G}\frac{l(\varphi_{+}(0)-\varphi_{-}(0))}{2}\\
&=\frac{1}{4G}\frac{i}{r\sqrt{1-n^2}}\left((l^2-r^2)\Pi\left[0;i\eta \Big|\frac{n^2(r^2-l^2)}{l^2(1-n^2)}\right]-(l^2-n^2r^2)\Pi \left[\frac{n^2r^2}{l^2};i\eta \Big|\frac{n^2(r^2-l^2)}{l^2(1-n^2)}\right]\right)\ .
\eal
\eeq
Using the identity of elliptic integral  
\beq
(p-q')\Pi[q';\phi |p]=p\Pi[0;\phi |p]-(p-q)\Pi[q;\phi|p]\ , \quad (1-q)(1-q')=1-p\ ,
\eeq
we can find it matches with the field theory result.

\section{Conclusions and Discussions}\label{cons}
In this paper, we focus on exploring the holographic dual of $T\bar T$ deformed BCFT. Based on the bottom-up AdS/BCFT duality and the cutoff description of holographic $T\bar T$ CFT, we propose that the bulk dual of $T\bar T$ BCFT is AdS gravity with both EOW brane and finite cutoff boundary, and the EOW brane intersects with the cutoff boundary at the $T\bar T$ BCFT boundary. We distinguish two Types of $T\bar T$ deformation by discussing the effect of the deformation on BCFT boundary. The first Type involves a deformed boundary, we calculate the boundary entropy to quantify the amount of boundary deformation. We also check the energy spectrum for a finite interval in this case. The second Type involves an undeformed boundary. We calculate the entanglement entropy
and Rényi entropy from both the field theory side and gravity side to provide evidence. 
Our study reveals a new realm of physics emerging from the combination of $T\bar T$ deformation and AdS/BCFT. This opens up an exciting avenue for investigating holographic CFT with multiple types of non-perturbative effect, including operator deformation, boundary etc.

There are some interesting questions for future exploration. First, in \cite{Guica:2019nzm}, the $T\bar T$ deformation is proposed to be dual to a mixed boundary condition on the asymptotic boundary. This proposal can address cases where matter is present in the bulk. Inspired by our proposal, it would be worthwhile to study holographic $T\bar T$ BCFT in terms of boundary conditions on the asymptotic boundary. Second, we can generalize AdS/$T\bar T$ BCFT to higher dimensions. Third, since the derivation of wedge holography is based on AdS/BCFT \cite{Akal:2020wfl}, it would be interesting to study $T\bar T$ deformation in wedge holography based on AdS/$T\bar T$ BCFT. Last but not least, it is interesting to compute other quantum information quantities under AdS/$T\bar T$ BCFT, such as reflected entropy and entanglement negativity.
\begin{acknowledgments}
We are grateful for the valuable discussions with Yang Zhou and the useful comments from Tadashi Takayanagi and Rongxin Miao.
\end{acknowledgments}

\newpage
\bibliographystyle{JHEP}
\bibliography{Holottbarb}

\end{document}